\documentstyle[prl,multicol,aps,epsfig,amssymb]{revtex}
 \tighten
\begin{document}

\draft

\title{Anomalous Low Temperature Behavior of Superconducting
Nd$_{\mathbf 1.85}$Ce$_{\mathbf 0.15}$CuO$_{\mathbf 4-y}$}

\author{L. Alff$^1$, S. Meyer$^1$, S. Kleefisch$^1$, U. Schoop$^1$,
A. Marx$^1$, H. Sato$^2$, M. Naito$^2$, and R. Gross$^1$}

\address{$^1$II.~Physikalisches Institut, Universit\"{a}t zu K\"{o}ln,
Z\"{u}lpicherstr.~77, D - 50937 K\"{o}ln, Germany}

\address{$^2$NTT Basic Research Laboratories, 3-1 Morinosato Wakamiya,
Atsugi-shi, Kanagawa 243, Japan}

\date{received June 29, 1999}
\maketitle
 \sf

\begin{abstract}

We have measured the temperature dependence of the in-plane London
penetration depth $\lambda_{ab}(T)$ and the maximum Josephson current
$I_c(T)$ using bicrystal grain boundary Josephson junctions of the
electron-doped cuprate superconductor Nd$_{1.85}$Ce$_{0.15}$CuO$_{4-y}$.
Both quantities reveal an anomalous temperature dependence below about
$4$\,K. In contrast to the usual monotonous decrease (increase) of
$\lambda_{ab}(T)$ ($I_c(T)$) with decreasing temperature, $\lambda_{ab}(T)$
and $I_c(T)$ are found to increase and decrease, respectively,  with
decreasing temperature below 4\,K resulting in a non-monotonous overall
temperature dependence. This anomalous behavior was found to be absent in
analogous measurements performed on Pr$_{1.85}$Ce$_{0.15}$CuO$_{4-y}$. From
this we conclude that the anomalous behavior of
Nd$_{1.85}$Ce$_{0.15}$CuO$_{4-y}$ is caused by the presence of the Nd$^{3+}$
paramagnetic moments. Correcting the measured $\lambda_{ab}(T)$ dependence
of Nd$_{1.85}$Ce$_{0.15}$CuO$_{4-y}$ for the temperature dependent
susceptibility due to the Nd moments, an exponential dependence is obtained
indicating isotropic $s$-wave pairing. This result is fully consistent with
the $\lambda_{ab}(T)$ dependence measured for
Pr$_{1.85}$Ce$_{0.15}$CuO$_{4-y}$.

\end{abstract}


\vspace*{-10cm}\noindent {\sc Physical Review Letters} \hfill accepted for
publication 9 August 1999
\vspace*{9.5cm}

\begin{multicols}{2}
\narrowtext

The vast majority of experiments on the cuprate superconductors are
performed on hole-doped materials. Much less attention has been paid to the
system $Ln_{2-x}$Ce$_x$CuO$_{4-y}$ (with $Ln=$ Pr, Nd, Sm, Eu)
\cite{Tokura:89} which represents an electron-doped material. Both hole- and
electron-doped cuprates have in common the copper oxygen planes as the
central building blocks of the high-temperature superconductors (HTS)
suggesting similar superconducting properties. However, as can already be
seen from the differences of the generic phase diagram on the electron- and
hole-doped side, the physics of electron- and hole-doped HTS is different.
In particular, the order parameter (OP) symmetry of the electron-doped
cuprates is most likely of $s$-wave type
\cite{Huang:90,Wu:93,Andreone:94,Alff:98}, in contrast to the $d$-wave OP
symmetry in the hole-doped HTS. To clarify the specific differences and
similarities between the electron- and hole-doped HTS a more detailed
experimental study of the electron-doped HTS is required.

Among the electron-doped materials, up to now Nd$_{2-x}$Ce$_{x}$CuO$_{4-y}$
has been the most intensively investigated material. This system is also
remarkable because of the significant influence of the magnetic moment of the
Nd$^{3+}$ ions, whereas in Pr$_{2-x}$Ce$_x$CuO$_{4-y}$ the Pr$^{3+}$ ion has a
singlet non-magnetic crystalline electric field ground state. It is well known
that the specific heat $C_p$ of Nd$_{2-x}$Ce$_{x}$CuO$_{4-y}$ shows a Schottky
anomaly at low temperatures and low Ce doping levels. This anomaly is
attributed to the splitting of the Nd-4$f$ ground-state doublet due to
interactions between the Nd moments and the ordered Cu moments
\cite{Brugger:93}. Surprisingly, a $C_p$ anomaly also was found for $x=0.15$,
even though no Cu ordering is observed for a Ce concentration above
$x\approx0.14$ \cite{Markert:89}. Compared to the Schottky anomaly at low Ce
doping levels this anomaly is shifted to lower temperatures and is
considerably broadened. Furthermore, the observed large value of the linear
specific heat coefficient $\gamma$ at temperatures below 1\,K is still
controversially discussed in terms of a novel type of heavy-fermion behavior
\cite{Fulde:93} and, alternatively, in terms of magnetic Nd excitations
\cite{Henggeler:97,Fuldecommentreply:99}. Nevertheless, it seems reasonable to
assume that at high Ce doping levels the Nd-Cu exchange is strongly reduced
and Nd-Nd interactions become more important inducing antiferromagnetic
coupling along the $c$-direction \cite{Henggeler:97}.

In this Letter, we report on the investigation of the influence of the Nd
magnetic moments on the superconducting properties of the fully oxygen reduced
compound Nd$_{1.85}$Ce$_{0.15}$CuO$_{4-y}$ (NCCO) with $x=0.15$ having a
maximum critical temperature $T_c\approx24$\,K. We measured the $T$ dependence
of the in-plane London penetration depth $\lambda_{ab}$, of the maximum
Josephson current $I_c$, as well as of the energy gap $\Delta$ using bicrystal
grain boundary Josephson junctions (GBJs). The advantage of the use of GBJs is
that all three measurements can be done using the same sample thereby
eliminating effects of different sample quality in the measurement of the
different quantities. In order to clearly establish the effect of the Nd
moments on the measured quantities comparative experiments have been performed
on Pr$_{1.85}$Ce$_{0.15}$CuO$_{4-y}$ (PCCO). The basic result of our
measurements is that for NCCO both $\lambda _{ab}(T)$ and $I_c(T)$ show
pronounced anomalies below about 4\,K, whereas such anomalies are absent for
PCCO. The $\lambda_{ab}(T)$ data of both NCCO, after being corrected for the
influence of the Nd moments, and PCCO is consistent with an isotropic $s$-wave
OP. For $4$\,K$<T<T_c$ the superconducting properties of NCCO and PCCO are
similar suggesting that at low $T$ the influence of the Nd$^{3+}$ moments
causes the difference between the two materials. The observation that $\Delta
(T)$ derived from tunneling spectra is almost identical for both materials
between 2\,K and $T_c$ indicates that the Nd moments are not coupled to the
superconducting electron system by conduction mediated processes. More likely,
the Nd-Nd interactions affect the superconducting properties of NCCO only
through relatively weak dipolar terms as already has been discussed e.~g.~in
the case of the hole-doped system GdBa$_2$Cu$_3$O$_{7-\delta}$
\cite{Dunlap:88}. The observation of a decreasing $I_C$ with decreasing $T$ in
NCCO Josephson junctions is a completely new effect that is believed to be
also associated with the Nd magnetic moments.

The NCCO- and PCCO-GBJs were fabricated by the deposition of $c$-axis oriented
NCCO and PCCO thin films on SrTiO$_3$ bicrystal substrates with misorientation
angles of $7^{\circ}$, $10^{\circ}$, and $24^{\circ}$ using molecular beam
epitaxy (MBE). A detailed description of the fabrication process was given by
Naito {\em et al.} \cite{Naito:95}. Josephson behavior in NCCO-GBJs has been
demonstrated by Kleefisch {\em et al.} \cite{Kleefisch:98}.  We stress that
the demonstration of the Josephson effect for both NCCO and PCCO, the low
resistivity values (below 50\,$\mu\Omega$cm at 25\,K) and the $T_c$ value of
about 24\,K demonstrate that the thin film samples are optimum doped, well
oxygen reduced, and single phase. This is important with respect to the
possibility of inhomogeneous dopant distribution or the formation of different
phases which is more difficult to control for large bulk single crystals.

\begin{figure}[tbh]
 \center{\includegraphics [width=1.0\columnwidth] {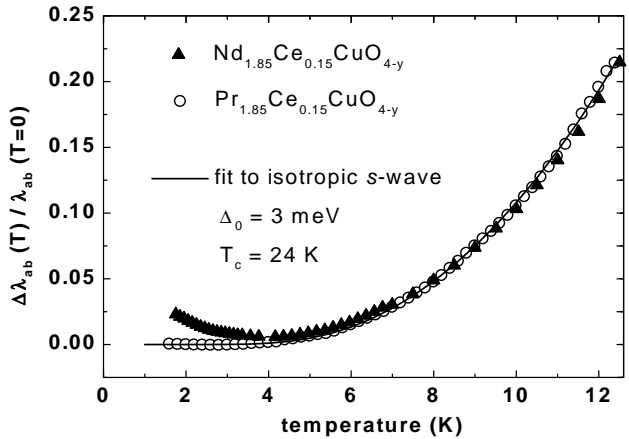}}
 \vspace*{-0.4cm}
\caption{Relative change $\Delta\lambda_{ab}$ of the London penetration
depth in NCCO and PCCO  as a function of temperature measured for a
symmetrical, 10$^{\circ}$ [001] tilt  GBJs.}
 \label{lambda}
\end{figure}

The maximum Josephson current $I_c$ of the GBJs was determined by standard
four probe technique in a magnetically shielded environment. The measurement
of $\Delta\lambda_{ab}$ is based on the measurement of $I_c$ as a function
of an applied magnetic field $H||c$. GBJs formed in different HTS have been
successfully used to determine the relative change
\[
\displaystyle\frac{\Delta\lambda_{ab}(T)}{\lambda_{ab}(T=0)} =
\displaystyle\frac{\lambda_{ab}(T) - \lambda_{ab}(T=0)}{\lambda_{ab}(T=0)}
\]
of the in-plane London penetration depth $\lambda_{ab}$ with the high
precision of below 1\,\AA. In this technique the shift of the side-lobes of
the $I_c(H)$ pattern of small Josephson junctions is measured as a function
of $T$. Details of this measurement technique have been described elsewhere
\cite{Froehlich:94}.

In Fig.~\ref{lambda} the relative change of $\lambda_{ab}$ is plotted versus
temperature. For PCCO, $\lambda_{ab}(T)$ behaves exponentially at low $T$
following $\Delta\lambda_{ab}(T)\propto \sqrt{\Delta /T}
\exp^{-\Delta/k_BT}$ as expected for a BCS-type isotropic $s$-wave
superconductor \cite{Muehlschlegel:59}. In contrast, for a $d$-wave
superconductor, the nodes in the OP cause a linear behavior
$\Delta\lambda_{ab}(T)\propto T/\Delta$ \cite{Annett:91}. This linear
behavior has been observed for hole doped HTS using the method described
above\cite{Froehlich:94}. A $d$-wave behavior clearly is not consistent with
the $\Delta\lambda_{ab}(T)$ data measured for PCCO. We note that
irrespective of the detailed OP symmetry, a {\em monotonous}
$\Delta\lambda_{ab}(T)$ dependence is expected in clear contrast with the
result obtained for NCCO below about 4\,K.

\begin{figure}[tbh]
 \center{\includegraphics [width=1.0\columnwidth] {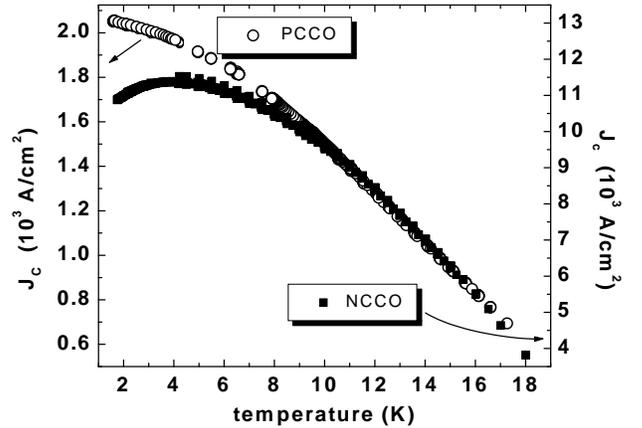}}
 \vspace*{-0.4cm}
\caption{Critical Josephson current density $J_c$ vs. temperature for
symmetrical, 10$^{\circ}$  [001] tilt NCCO- and PCCO-GBJs. } \label{ict}
\end{figure}

In Fig.~\ref{ict}, $J_c(T)$ is plotted for NCCO- and PCCO-GBJs. Whereas a
monotonous increase of $I_c(T)$ with decreasing $T$ is found for PCCO, for
NCCO $I_c$ is found to decrease again at low $T$. A monotonous $I_c(T)$
dependence as found for PCCO-GBJs is also observed for GBJs fabricated from
the hole doped HTS.

\begin{figure}[tbh]
 \center{\includegraphics [width=0.95\columnwidth] {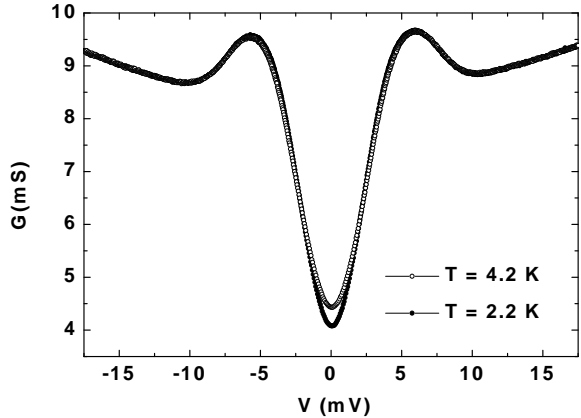}}
 \vspace*{-0.1cm}
\caption{Tunneling conductance vs. voltage of a  symmetrical, 24$^{\circ}$
[001] tilt  NCCO-GBJ at 4.2\,K and 2.2\,K.}
 \label{gap}
\end{figure}

In order to interpret our NCCO data, at first sight it is tempting to assume a
reduction of the superconducting OP with decreasing $T$ due to enhanced
pair-beaking by magnetic scattering e.~g.~similar to the ternary molybdenum
chalcogenides (Chevrel phases) \cite{Fischer:78}. Then, a reduced superfluid
density $n_s$ could be the origin of the observed anomalies because of
$J_c\propto n_s$ and $\lambda_{ab}(T)\propto1/\sqrt{n_s}$. However, in
tunneling measurements performed on the same samples no significant change of
the gap value is observed at low $T$ both for NCCO and PCCO. Moreover, the
tunneling conductance $G$ at zero bias still decreases when $T$ is lowered
from 4\,K to 2\,K as shown in Fig.~\ref{gap}. These experimental facts give
strong evidence against the assumption of reduced $n_s$ or $\Delta$ as the
origin of the observed anomalies.

Next we briefly discuss whether the observed anomalies in $\lambda_{ab}(T)$
and $I_c(T)$ can be caused by a $d$-wave OP. It has been both theoretically
predicted and experimentally observed that $\lambda_{ab}(T)$ can increase with
decreasing $T$ due to so-called anti-Meissner currents related to Andreev
bound states \cite{Alff:epj,Walter:98}. However, assuming
$d_{x^2-y^2}$-symmetry of the OP the maximum spectral weight of such bound
states is expected for interfaces close to the (110) orientation. As a
consequence this effect has only been observed in large angle grain boundaries
\cite{Alff:epj}, whereas the $T$ dependence of $\Delta\lambda_{ab}(T)$ for
small misorientation angles ($\theta\lesssim 24^{\circ}$) was found to behave
linearly as expected for a $d$-wave OP\cite{Froehlich:94}. Furthermore, the
YBCO-GBJs showing this kind of anomaly in $\lambda_{ab}(T)$ also show a
$I_c(H)$ pattern that strongly deviates from a Fraunhofer pattern and has the
maximum $I_c$ value shifted to $H\neq0$. This indicates the presence of
negative currents due to the $d$-wave symmetry of the OP in the electrodes
\cite{Mannhart:96}. In contrast, the $I_c(H)$ patterns of NCCO- and PCCO-GBJs
are close to a regular Fraunhofer diffraction pattern expected for an ideal
Josephson junction \cite{Kleefisch:98}. Finally, the tunneling spectra
measured for NCCO do not show any zero bias anomaly giving strong evidence for
the absence of a sign change in the superconducting pair potential
\cite{Alff:98,Alff:epj,Kashiwaya:98}. These experimental facts exclude both a
$d_{x^2-y^2}$- and $d_{xy}$-symmetry of the OP for NCCO and PCCO.

\begin{figure}[tbh]
 \center{\includegraphics [width=1.0\columnwidth] {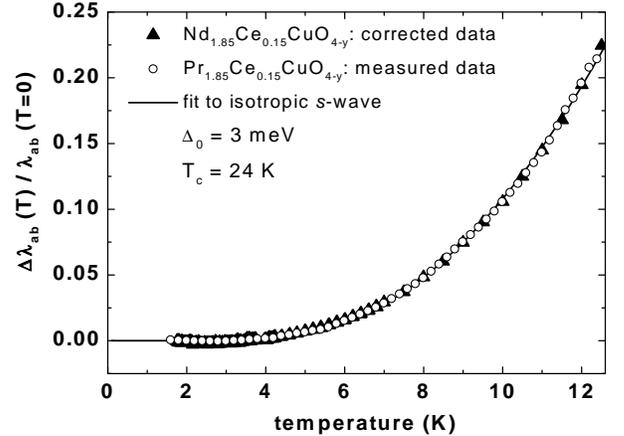}}
 \vspace*{-0.4cm}
\caption{$\Delta\lambda_{ab}(T)/\lambda_{ab}(0)$ for NCCO and PCCO. The NCCO
data is corrected as described in the text.}
 \label{corrected}
\end{figure}

We now discuss the influence of the Nd magnetic moments on the measured $T$
dependence of the London penetration depth of NCCO. The anomaly in the
$\Delta\lambda_{ab}(T)$ dependence of NCCO can be understood by taking into
account the effective magnetic moment of the Nd$^{3+}$ ions as recently
proposed by Cooper \cite{Cooper:96}. The measured paramagnetic susceptibility
$\chi$ of NCCO with the magnetic field applied parallel to the $c$-axis fits
to a Curie-Weiss law $\chi(T)=\chi_0+C/(T+\Theta)$, where $C$ is the Curie
constant and $\Theta$ the Curie-Weiss temperature\cite{Dali:93}. This results
in a magnetic permeability $\mu(T)=1+4\pi\chi(T)\approx 1+const./(T+\Theta)$.
In our analysis we assume that this dependence holds down to $T\approx2$\,K
with $C\approx0.2$\,emu\,K/mole Nd (corresponding to an effective moment of
about 1.2\,$\mu_B$ per Nd$^{3+}$ ion). The effect of the strong $T$ dependence
of the magnetic susceptibility on $\lambda_{ab}(T)$ is as follows: On the one
hand, the solution of London's equation shows that the measured penetration
depth has to be multiplied by the additional factor $\sqrt{\mu(T)}$ in order
to reveal the London penetration depth mirroring the superfluid density $n_s$.
On the other hand, for the maximum Josephson current the flux density
${\mathbf{B}}=\mu(T)\mu_0{\mathbf{H}}$ threading the GBJ is relevant. Hence,
in total, for our experimental method the measured penetration depth has to be
divided by $\sqrt{\mu(T)}$ to reveal the intrinsic penetration depth
reflecting $n_s$. As can be seen from Fig.~\ref{corrected}, the anomaly in
$\lambda_{ab}(T)$ can be entirely attributed to the influence of the $T$
dependent permeability due to the Nd moments. After correcting the NCCO data
they coincide with the PCCO data supporting the validity of the data
correction.  The intrinsic penetration depth derived in this way can be well
fitted to the exponential $T$ dependence of an isotropic $s$-wave
superconductor. Assuming a $d$-wave OP no reasonable fit could be obtained. We
note that this result is in clear contrast to the recent conclusion of Cooper
who suggested that $\lambda_{ab}(T)$ of NCCO might be consistent with a
$d$-wave OP after data correction. In contrast, our data for both NCCO and
PCCO is clearly consistent with an isotropic $s$-wave symmetry of the OP. We
further note that there is still a discrepancy between tunneling measurements
suggesting an {\em an\/}isotropic $s$-wave OP for NCCO
\cite{Alff:epj,Kashiwaya:98} and the present $\Delta\lambda_{ab}(T)$
measurements indicating an almost isotropic $s$-wave OP for both NCCO and
PCCO. However, both measurements are clearly {\em not} consistent with a
$d$-wave OP in the electron-doped HTS.

We finally discuss the anomalous behavior of $I_c(T)$ for NCCO-GBJs. As shown
in Fig.~\ref{ict},  $I_c$ is reduced by about 15\% at 2\,K as compared to the
value extrapolated from the $I_c(T)$ dependence at higher $T$. An equivalent
reduction would be obtained by applying a magnetic field of about $0.2$\,G.
Since the anomalous $I_c(T)$ dependence is absent for the PCCO-GBJs, it is
natural to attribute this effect also to the presence of the Nd magnetic
moments. However, at present we have no conclusive theoretical understanding
of the origin of the $I_c(T)$ anomaly. From neutron scattering experiments in
NCCO with high Ce doping levels it has been concluded that a short-range order
of the diluted Nd moments is gradually established over a large temperature
range with no clearly defined N\'{e}el temperature \cite{Henggeler:97}.
Unfortunately, until now it is unclear whether the NCCO samples used for
neutron scattering were fully oxygen reduced and whether a possible Nd
ordering can create a small effective magnetic field reducing $I_c$. Another
possibility is the creation of an effective field at the grain boundary
interface due to disorder.  To clarify the detailed origin of the anomalous
$I_c(T)$ dependence, experiments directly probing the magnetic interactions in
the optimum superconducting compound are desirable. Also, the possible
presence of magnetic flux in the grain boundary region might be checked by
comparative scanning SQUID measurements on NCCO- and PCCO-GBJs. We finally
note that our measurement can not exclude the possibility of a coupling of the
Nd spin system to the conduction electrons at still lower temperatures
($T<1$\,K) \cite{Fulde:93}. We also would like to point out that anomalous
$\Delta\lambda_{ab}(T)$ and $I_c(T)$ dependencies are expected for magnetic
rare-earth substituted hole-doped HTS \cite{Ormeno:pre}. However, in contrast
to NCCO the corrected $\Delta\lambda_{ab}(T)$ data are expected to be
consistent with a $d_{x^2-y^2}$-wave symmetry of the OP.

As a consequence of the anomalous low $T$ behavior of NCCO it is evident that
Nd$_{2-x}$Ce$_x$CuO$_{4-y}$ is not well suited for a detailed comparison of
the superconducting properties of electron-doped to the corresponding
hole-doped cuprate superconductors. This is in particular important for
measurements of $\lambda_{ab}(T)$ and conclusions drawn from such measurements
with respect to the symmetry of the OP. In this context some recent
measurements have to be re-interpreted. For comparative measurements of the
penetration depth of hole- and electron-doped HTS at $T< 4$\,K, the Pr doped
compound certainly is the better choice. However, there is no problem with the
results of tunneling measurements. It has been shown that there is no
qualitative difference in the tunneling spectra between NCCO and PCCO for
$2$\,K$<T<T_c$ \cite{Alff:98,Alff:epj}.

In conclusion, we have observed an anomalous low temperature dependence of the
in-plane London penetration depth and the maximum Josephson current of
Nd$_{1.85}$Ce$_{0.15}$CuO$_{4-y}$-GBJs. The absence of the anomalous behavior
in Pr$_{1.85}$Ce$_{0.15}$CuO$_{4-y}$-GBJs strongly suggests that the anomalies
observed for NCCO are caused by the Nd$^{3+}$ magnetic moments. The anomalous
$\lambda_{ab}(T)$ dependence is in good agreement with theoretical predictions
based on a Curie-Weiss type temperature dependence of the magnetic
susceptibility of NCCO affecting the measurement of $\lambda_{ab}(T)$. A
possible $d$-wave order parameter symmetry in NCCO that also could account for
an anomalous behavior can be excluded. In contrast, our $\lambda_{ab}(T)$ data
are consistent with an isotropic $s$-wave order parameter for both
investigated electron-doped HTS.

This work is supported by the Deutsche Forschungsgemeinschaft (SFB 341). The
authors acknowledge helpful discussions with S.~Anlage, H.~Berg,
B.~B\"{u}chner, M.~Fogelstr\"{o}m, M.~R\"{o}pke, and B.~Roessli.

\end{multicols}
\end{document}